\documentclass[fleqn,usenatbib]{mnras}

\usepackage{newtxtext,newtxmath}

\usepackage[T1]{fontenc}
\usepackage{ae,aecompl}

\usepackage{graphicx}	
\usepackage{amsmath}	
\usepackage{amssymb}	

\def\deg {$^{\circ}$}
\newcommand{\HI}{H$\,${\sc i}}
\newcommand{\HII}{H$\,${\sc ii}}

\newcommand{\dcm}{cm$^{-3}$}

\def\ts   {\thinspace}
\def\kms  {km\,s$^{-1}$}
\def\mo   {\ifmmode{{\rm M}_{\odot}}\else{M$_{\odot}$}\fi}

\def\bco {\ifmmode{^{12}{\rm CO}(J=2\to1)}\else{$^{12}{\rm
CO}(J=2\to1)$}\fi}
\def\m  {\ifmmode{\mu {\rm m}}\else{$\mu$m}\fi}
\def\cco {\ifmmode{^{13}{\rm CO}(J=1\to0)}\else{$^{13}{\rm
CO}(J=1\to0)$}\fi}
\def\dco {\ifmmode{^{13}{\rm CO}(J=2\to1)}\else{$^{13}{\rm
CO}(J=2\to1)$}\fi}
\def\eco {\ifmmode{{\rm C}^{18}{\rm O}(J=1\to0)}\else{{\rm C}$^{18}{\rm
O}(J=1\to0)$}\fi}
\def\hi  {{H\ts {\scriptsize I}}\fi}
\def\hii  {{H\ts {\scriptsize II}}}

\def\Hb  {\ifmmode{{\rm H}{\alpha}}\else{H\ts {$\beta$}}\fi}

\def\nh  {\ifmmode{N(\hi)}\else{$N$(\hi)}\fi}
\def\hun  {\ifmmode{I_{100}}\else{$I_{100}$}\fi}
\def\sex  {\ifmmode{I_{60}}\else{$I_{60}$}\fi}
\def\hh   {\ifmmode{{\rm H}_2}\else{H$_2$}\fi}
\def\nhh   {\ifmmode{N({\rm H}_2)}\else{$N$(H$_2$)}\fi}
\def\zwco  {\ifmmode{^{12}{\rm CO}}\else{$^{12}{\rm CO}$}\fi}
\def\nzwco  {\ifmmode{N(^{12}{\rm CO})}\else{$N(^{12}{\rm CO})$}\fi}
\def\wzwco  {\ifmmode{W(^{12}{\rm CO})}\else{$W(^{12}{\rm CO})$}\fi}
\def\drco  {\ifmmode{^{13}{\rm CO}}\else{$^{13}{\rm CO}$}\fi}
\def\ndrco  {\ifmmode{N(^{13}{\rm CO})}\else{$N(^{13}{\rm CO})$}\fi}
\def\wdrco  {\ifmmode{W(^{13}{\rm CO})}\else{$W(^{13}{\rm CO})$}\fi}
\def\tex  {\ifmmode{T_{ex}({\rm CO})}\else{$T_{ex}({\rm CO})$}\fi}
\def\xco   {\ifmmode{X_{\rm CO}}\else{$X_{\rm CO}$}\fi}
\def\msol   {\ifmmode{{\rm M}_{\odot}}\else{M$_{\odot}$}\fi}

\def\water {H$_2$O} 

\def\Lsun{L$_{\odot}$}

\title[Evolved Star Atmospheres and Milky Way Spiral Arms]{Evidence for Coupling of Evolved Star Atmospheres and Spiral Arms of the Milky Way}

\author[Gorski \& Barmby]{
Mark D. Gorski,$^{1}$\thanks{E-mail: mgorski3@uwo.ca},
Pauline Barmby,$^{1,2}$\thanks{E-mail: pbarmby@uwo.ca},
\\
$^{1}$Department of Physics \& Astronomy,
$^{2}$Institute for Earth \& Space Exploration, University of Western Ontario, 1151 Richmond Street, London, Ontario, N6A 3K7, Canada\\
}

\date{Accepted XXX. Received YYY; in original form ZZZ}

\pubyear{2019}

\begin{document}
\label{firstpage}
\pagerange{\pageref{firstpage}--\pageref{lastpage}}
\maketitle

\begin{abstract}
It is imperative to map the strength and distribution of feedback in galaxies to understand how feedback affects the galactic ecosystems. 
\water\ masers act as indicators of energy injection into the ISM.
Our goal is to measure the strength and distribution of feedback traced by water masers in the Milky Way. 
We identify optical counterparts to \water\ masers discovered by the HOPS survey.
The distribution and luminosities of \water\ masers in the Milky Way are determined using parallax measurements derived from the second Gaia Data Release. 
We provide evidence of a correlation between evolved stars, as traced by \water\ masers, and the spiral structure of the Milky Way, suggesting a link between evolved stars and the Galactic environment. 
\end{abstract}



\section{Introduction}



It is imperative to understand the effect of feedback on galaxy ecosystems. 
\citet{Kennicutt2007,Leroy2013,Meidt2015} have shown that the star formation efficiency throughout galaxies changes depending on the galactic environment. 
In addition simulations show that without feedback galaxies' gas stores would collapse into stars in less than a dynamical time \citep[e.g.][]{Kauffmann1999,Hopkins2011,Krumholz2011}.
Furthermore feedback effects are predicted to combine non-linearly \citep{Hopkins2012,Hopkins2014} with observational evidence recently provided by \citet{Gorski2017, Gorski2018}.
Thus understanding the strength and distribution of stellar feedback within galaxies is a critical step in understanding how galaxy ecosystems are regulated.

Many astrophysical environments give rise to 22.235~GHz o-\water\ ($6_{1,6}-5_{2,3}$) maser  emission. 
These masers are found in high mass star forming regions \citep[e.g.,][]{Genzel1977,Urquhart2009,Urquhart2011}, low mass protostars \citep[e.g.][]{Dickinson1974,Furuya2003,Furuya2007a,Furuya2007b}, planetary nebulae \citep{Miranda2001}, and active Galactic nuclei (AGN)  \citep[e.g.,][]{Herrnstien1999,Reid2009}.
\citet{Cesaroni1988} notes water masers associated with T Tauri or Herbig Haro objects. 
\citet{Hagiwara2001} describes the classification of \water\ masers in the literature: stellar \water\ masers have typical luminosities of $\sim10^{-3}$~\Lsun, kilomasers have lumiosities of $0.1-1$~\Lsun, and megamasers have luminosities greater than 20~\Lsun.

The \water\ molecule requires dense gas (10$^6$~\dcm) and high temperatures (T$\sim$400~K) to mase \citep{Elitzur1989}.
Water masers thus trace shocked gas in the environments around young or evolved stars, and AGN. 
For stellar \water\ masers, \citet{Engels1986} showed that there is a correlation between the mass loss rate of the parent star and the luminosity of the \water\ maser. 
Despite the different progenitors of the maser emission, the detection of a \water\ maser appears to be an indicator of feedback into the interstellar medium~(ISM). 
That is to say, for stellar \water\ masers, energy and momentum is injected into the ISM at the beginning and end of the stellar life cycle.

Prior to this work the most recent attempt to understand the statistical distribution of stellar \water\ masers was by  \citet{Palagi1993}. 
They compared the Arcetri atlas of \water\ maser sources \citep{Comoretto1990} with the IRAS point source catalogue.
The angular resolution of the Arcetri atlas was poor by contemporary standards ($\sim$2\arcmin).
Distances were determined using the kinematic distance method with a rotation curve from \citet{Brand1986}.
Due to the uncertainty in maser site velocities, distance uncertainties were often very high.
For star formation masers \citet{Palagi1993} limited their analysis to sources  with distance uncertainties less than 50\%, reducing the 224 sources to 181. 
Additionally \water\ maser velocity may not accurately trace the kinematics of their parent system, with the largest known offset in the Milky Way being 100 \kms{} \citep{Titmarsh2013}.

In this work we compare the 22~GHz \water\ maser sources derived from the \water\ southern Galactic Plane Survey \citep[HOPS:][]{Walsh2014} and Gaia Data Release 2 \citep{GaiaDR22018}. 
By using the Gaia DR2 data set we avoid the large uncertainties involved with estimating kinematic distances to star formation masers.
With an order of magnitude increase in  spatial resolution, we investigate the distribution and three dimensional positions of stellar water masers in the Milky Way using Gaia derived distances.
Care must be taken as the Gaia data were obtained in optical wavelengths and dust obscures much of the Galactic plane.
By comparison, radio wavelengths are largely unaffected by dust.  
The goal of this project is to yield the distribution and strength of stellar feedback in the Milky Way by measuring the luminosity and position of \water\ masers within the Galaxy.

In \S \ref{Methods} we describe the data sets used for our analysis and the methods for combining the two catalogues and classification of the optical counterparts. 
In \S \ref{maser dist} we compare the observed distribution of masers with Galactic structure. 
In \S \ref{Discussion} we discuss the results and conclusions are presented in \S \ref{Conclustions}.

\section{Methods}
\label{Methods}

\subsection{HOPS and Gaia DR2 data}

HOPS \citep{Walsh2011}  is the largest unbiased \water\ maser survey to date.
The survey covers 100 square degrees of sky between Galactic longitudes~$290^{\circ}<\ell< 90^{\circ}$ and Galactic latitudes~$-0.5^{\circ} <b<+0.5^{\circ}$. 
The original survey was observed with the Mopra radio telescope with a flux density limit of $\sim$1--2~Jy, with a gap at 10$^{13}$~W~Hz$^{-1}$, and a velocity resolution of 0.52~\kms. 
The angular resolution of the Mopra telescope is poor ($\sim2$\arcmin), so observations to better constrain maser positions were made with the Australia Telescope Compact Array,  with beam sizes ranging from $0.55\times0.35$\arcsec\ to $14.0\times10.2$\arcsec\ \citep{Walsh2014}.
The typical positional accuracy is about 1\arcsec. 
The observations have rms noise levels ranging from 6.5~mJy to 1.7~Jy; 90\% of the noise levels are in the 15--167~mJy range with the  peak of the distribution at 17~mJy.
\citet{Walsh2014} defined the terminology of observing \water\ masers with high spatial resolution ($\lesssim$1\arcsec) in the Milky Way. 
A maser site is a location on the sky subtending a spatial extent of order $\sim$1\arcsec.
Each maser site can be made up of many maser spots.
A maser spot is an isolated position with a single peak in the maser spectrum.
\citet{Walsh2014} detected 2790 maser spots consisting of 631 maser sites.

\begin{figure}
\centering
\includegraphics[width=0.45\textwidth]{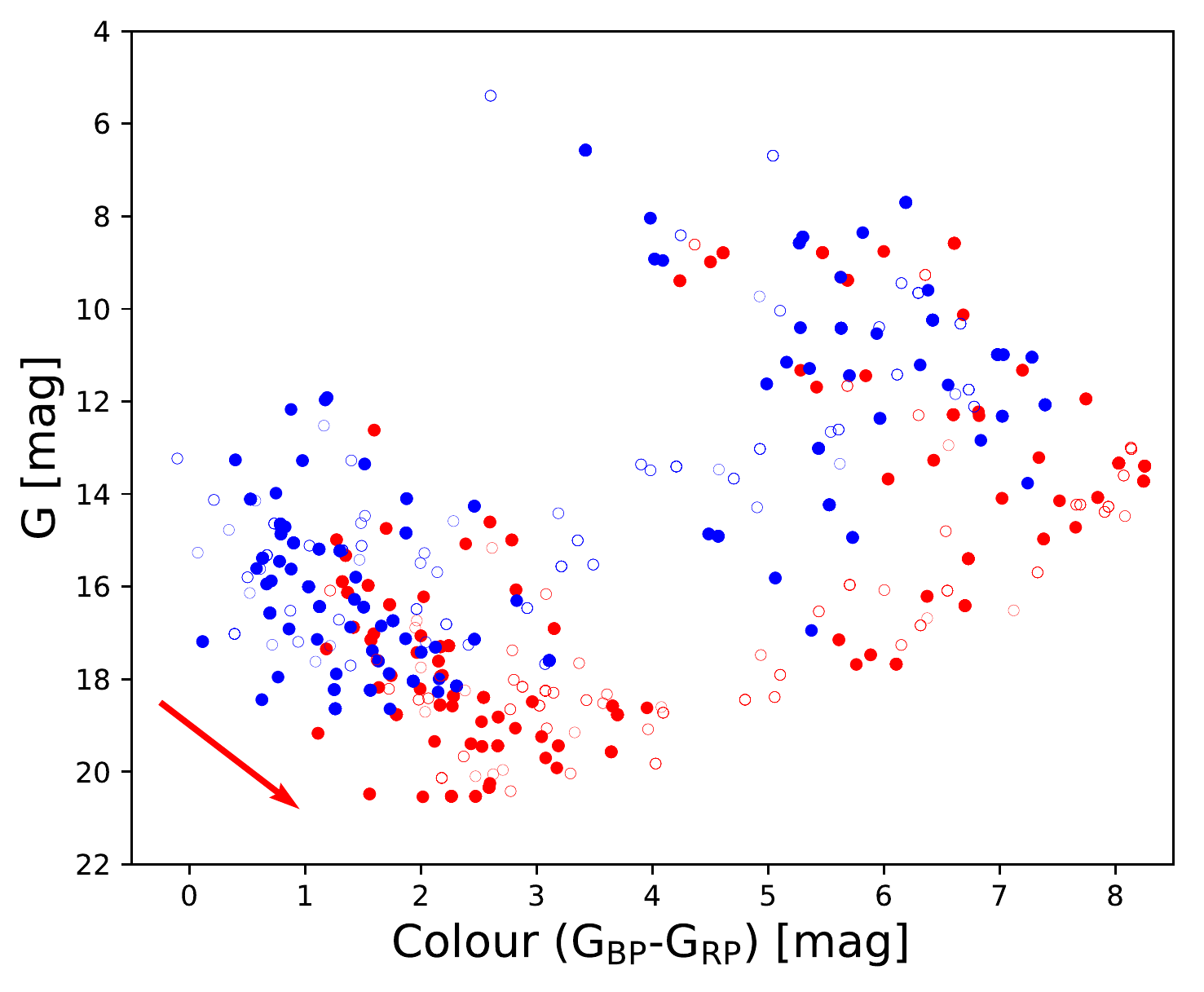}
\caption{
Colour magnitude diagram of optical counterparts to Milky-Way \water\ masers. 
Red circles show the observed magnitudes and blue filled circles show the extinction corrected magnitudes using the extinction maps from \citet{Chen2017}.
Open circles show sources that have been insufficiently extinction corrected because  The source's derived distance is greater than the maximum reliable distance from \citet{Chen2017}.
The average reddening vector (A$_{\rm G}$=2.3) is shown as an arrow in the bottom left.}
\label{fig:cmd}
\end{figure}

With the second Gaia Data Release \citep[DR2;][]{GaiaDR22018} we can match the distances to stellar \water\ masers with their Milky Way counterparts.
This data release provides parallax measurements for $\sim1.3$ billion stars and photometric measurements in G band (350-1000~nm) towards $\sim1.6$ billion sources with a limiting G band magnitude of 21.
It also includes blue (G$_{bp}$; 380-680~nm) and red (G$_{rp}$: 640-1000~nm) photometric measurements towards $\sim1.3$ billion sources.
After identifying Gaia sources associated with stellar \water\ masers,
we can measure the spatial distribution and properties of \water\ masers throughout the Milky Way and the distance-dependent properties of their progenitor sources. 

\subsection{Gaia Distances}

In order to determine the distance to each maser site we matched the Gaia ID to the catalogue of \citet{BailerJones2018}.
Inverting the parallax of the individual optical counterparts is not a reliable method to determine the distance to Gaia sources. 
It is best to use a probabilistic approach. 
The details are described by \citet{BailerJones2018} and the references therein.
Distances are inferred with a Bayesian approach using an exponentially decreasing space density as a prior. 
This does not make any assumptions about the  properties of individual stars or the extinction along their line of sight.
The uncertainties we use in the distance determination  are the 1$\sigma$ bounds of the asymmetric probability density function.

\subsection{Cross Referencing}

Maser sites are associated with a single astrophysical object and can contain many individual spots. 
\citet{Forster1989} studied the separation between \water\ maser emission and continuum emission from \HII\ regions. 
They found that the median separation is 20~mpc (0.65~ly), though separations can be as large as 130~mpc. 
The median separation distance corresponds to  angular separations of 0.5\arcsec\ at a distance of 8.5~kpc and 4.1\arcsec\ at a distance of 1~kpc. 
These angular separations are consistent with the findings by \citet{Walsh2014} that water masers cluster in sites  with diameters less than four arcseconds. 
Searching for stellar counterparts for maser sites, we found the average position of each maser site by calculating the mean right ascension and declination of all spots contained with a maser site. 
The best match to this calculated position within four arcseconds was found. 
This radius was chosen as the angular distribution of  \water\ masers on the sky dominates the ability to determine optical counterparts.

It was appropriate to search for near-infrared counterparts before looking for optical counterparts,
because optical bands are heavily affected by dust.
We first matched the average centre of each maser site with 2MASS \citep{Skrutskie2006} sources within a 4\arcsec\ radius using the Topcat software \citep{Taylor2005}.
We use 2MASS because it provides greater astrometric precision than WISE \citep{Pihlstrom2018}.
Next we found the most likely 2MASS/Gaia cross-matched counterpart from \citet{Marrese2019}.
The 2MASS Survey has an effective angular resolution of 2.5\arcsec\ \citep{Skrutskie2006, Marrese2019} whereas the angular resolution of Gaia DR2 is $\sim$0.4\arcsec\ \citep{GaiaDR22018}.
There is thus ample opportunity for multiple Gaia sources to be matched to a single 2MASS source.
A sample of a dense field near the galactic plane from \citet{Arenou2018} measures $\sim450000$ Gaia sources per square degree or $\sim1.7$ Gaia sources in a 4\arcsec\ search radius.
In a sparse region \citet{Arenou2018} measures $\sim250000$ Gaia sources per square degree or $\sim0.9$ Gaia sources in our 4\arcsec\ search radius.

Gaia counterparts to the 2MASS sources are divided into two categories: good neighbours and best neighbour. Good neighbours are sources with compatible positions within the uncertainties of the two catalogues. Overall, in the Gaia and 2MASS catalogues, there are at most three neighbours and 99.78\% of sources have only one neighbour. If more than one neighbour is found then the best neighbour is chosen from the Figure of Merit (FoM; \citealp{Marrese2019}). 
The FoM is calculated from the angular distance between the sources, the position uncertainties, the epoch difference, and local source density using a Poisson distribution.
The exact form of the FoM is \citep{Marrese2017}:
\begin{equation}
    FoM(r)=\frac{1}{2\pi\rho^\prime}\exp{\Big(-\frac{1}{2}r^2\Big)}
\end{equation}
where $r$ is the distance and $\rho$ is the local surface density of sources.
$\rho$ depends on the initial search radius which is defined as 60\arcsec\ in addition to a radius needed to include the cross match output of stars with proper motions of 0.2\arcsec~yr$^{-1}$. This is to account for epoch differences between catalogues \citep{Marrese2017}.
Of the 2790 maser spots (631 sites) we find 2MASS counterparts towards 1605 spots (350 sites).

\begin{figure*}
    \centering
    \includegraphics[width=0.9\textwidth]{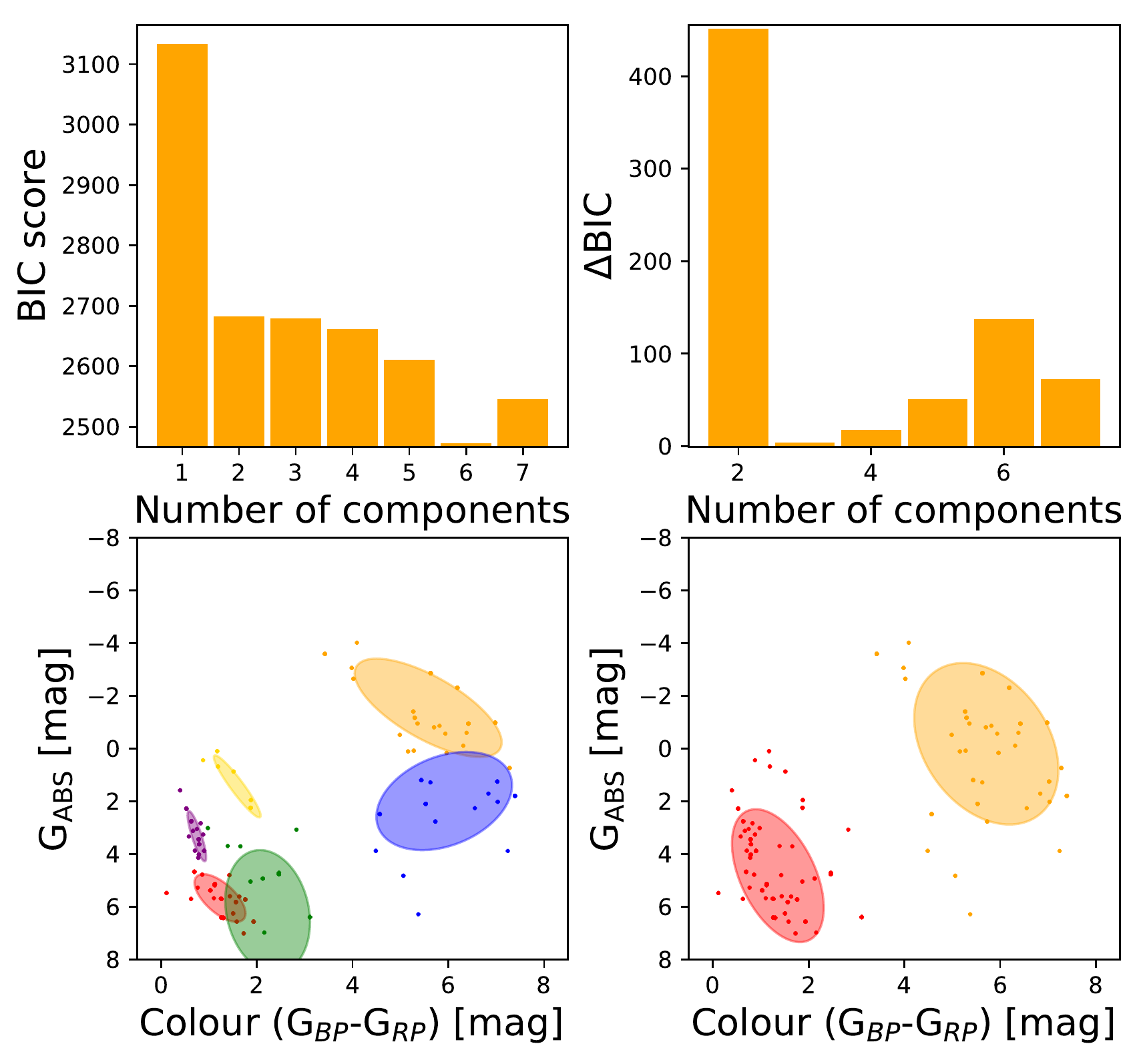}
    \caption{BIC scores and models. The left panels show the BIC score of models with up to seven components fit to the colour magnitude diagram. The model with the lowest BIC score is shown on the bottom left. The right panels show the change in BIC with the best fit to the data shown on the bottom right. }
    \label{fig:BIC}
\end{figure*}

\subsection{Extinction Correction}

The observed colour magnitude diagram for Gaia sources matched to HOPS \water\ maser spots is shown in Figure~\ref{fig:cmd}.
Interstellar extinction can have drastic effects on observed source properties \citep[e.g.,][]{Chen2017}.
The classification of stellar water masers from the optical counterparts will likely depend on how well we account for extinction and reddening in the Galactic plane.
We use three dimensional maps of interstellar dust from \citet{Chen2019} to correct for Galactic extinction. 
The maps were created using a machine-learning algorithm called Random Forest regression. 
We chose this catalogue as it covers the entire Galactic plane ($0^{\circ}<\ell <360^{\circ}$) and latitudes between $-10^{\circ}<b<10^{\circ}$, covering the entire survey area of HOPS. 
Furthermore, the catalogue provides colour excesses  ($E(G-K_{\rm S})$, $E(G_{\rm BP}-G_{\rm RP})$, and $E(H-K_{\rm S})$) which are converted into line-of-sight extinctions in the Gaia DR2 bands (see equations 3, 5, and 6 of \citealp{Chen2019}). The resolution of the maps is 6\arcmin\ out to a maximum distance of 6~kpc from the Sun. 
Figure \ref{fig:cmd} shows the extinction-corrected colour and apparent magnitudes of optical counterparts.
For sources that have distance estimates greater than 6~kpc we apply the value at 6~kpc. 
The average G magnitude of a matched maser counterpart is 14.3~mag and the average corrected $G_{\rm BP}-G_{\rm RP}$ colour is 3.0~mag.
In Figure \ref{fig:cmd} open circles show sources with distances greater than 6~kpc.
Sources further than 6~kpc could not be appropriately corrected for extinction. 

\subsection{Classification of Stellar Counterparts}
\subsubsection{Gaussian Mixture Modelling}

\citet{Palagi1993} classified stellar class water masers into two groups. 
They named these two groups SFR and STAR for masers associated with star forming regions or late type stars, respectively.
\citet{Walsh2014} classifies masers into three categories: star formation, evolved star, and other (unknown). 
When dealing with a large unclassified survey it is difficult to know how many classifications best describe the data. 
In order to separate the populations of stellar class \water\ masers we impose a Gaussian mixture model with an unspecified number of components to investigate how many groups best describe stellar counterparts.

We use the Bayesian Information Criterion (BIC) to determine the number of groups associated with stellar counterparts. 
Generally the lowest BIC score indicates the best fit model to the data. 
However one can fit an arbitrarily large number of components to the data almost always yielding a better fit than a fit using fewer components. 
The change in BIC ($\Delta$BIC) can be used to indicate the number of groups that best describe the data.  
We fit up to seven  groups to the colour magnitude diagram using a Gaussian mixture model from \citet{scikit-learn};
increasing the number of groups beyond seven did not result in improved fits.
Figure \ref{fig:BIC} shows Gaussian mixture models fit to the colour magnitude diagram of \water\ maser counterparts. 
The best fit to the data consists of six groups, however two groups best describe the data as indicated by the largest change in BIC, 
$\Delta$BIC=451 transitioning from one to two groups (Figure \ref{fig:BIC}).

\begin{figure}
    \centering
    \includegraphics[width=0.45\textwidth]{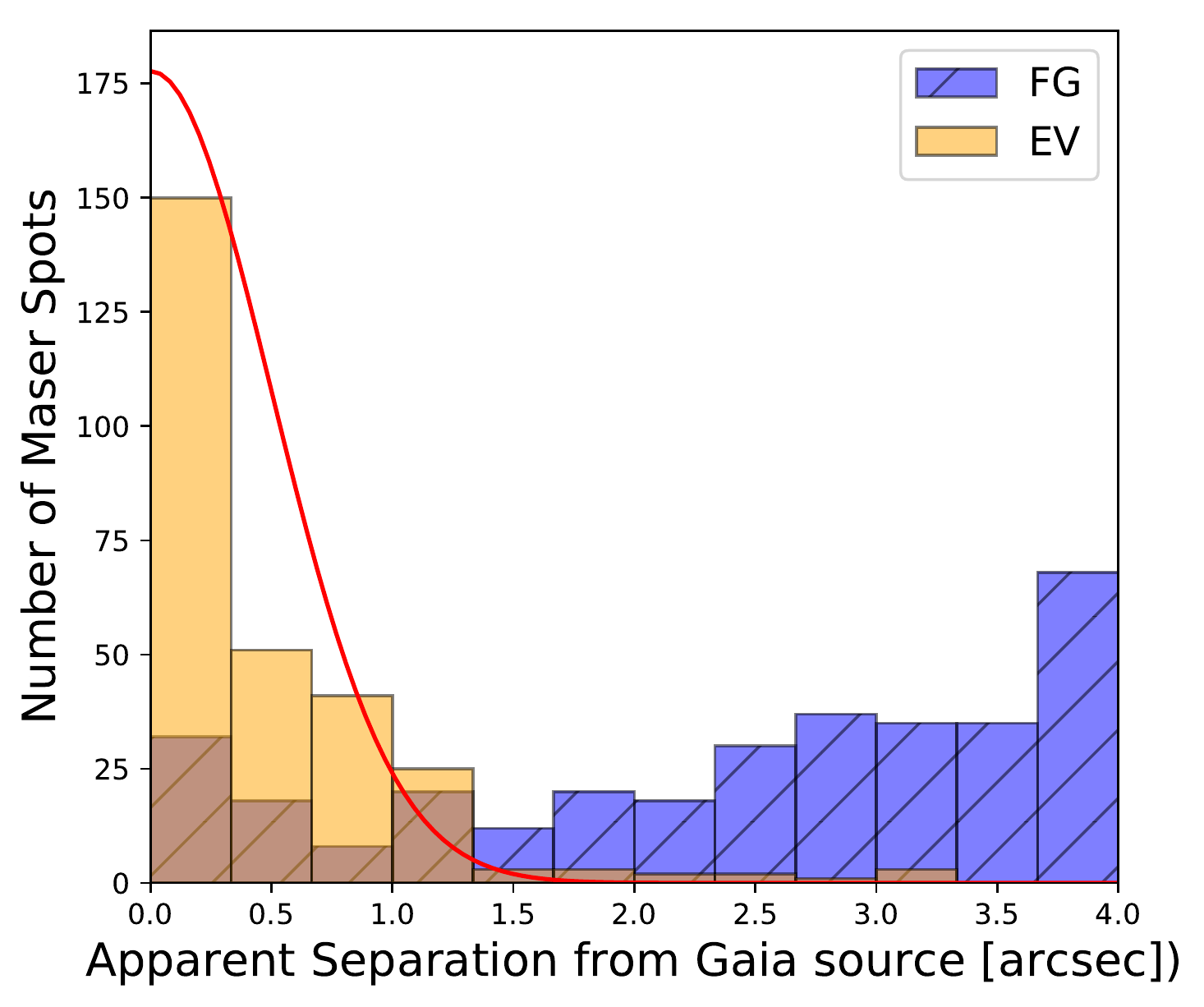}
    \caption{The number of maser spots and their  apparent angular distance from their Gaia counterpart. 
    The red line show a Gaussian fit to the evolved group distribution with a FWHM of $1.17\pm0.11$\arcsec.
    }
    \label{fig:offset_N}
\end{figure}

Identifying two groups in colour-magnitude space does not describe the physics of the population, so we must determine what these groups signify. 
The two groups of stellar counterparts populate the high-luminosity red quadrant of the colour magnitude diagram and lower-luminosity blue quadrant.
If the masers are related to star formation,  an \HII\ region should be apparent unless the the star formation is obscured. 
We expect stellar matches to be spatially coincident with their parent star on the sky.
Upon visual inspection of the resulting matches with DSS images, we find that Gaia-maser matches in the lower luminosity blue quadrant of the colour magnitude diagram do not have \HII\ regions within 1\arcmin\ around the Gaia source.
This is unsurprising as \HII\ regions that generate water masers are often deeply embedded with A$_v \gtrsim 20$ \citep[e.g.,][]{Moore1988, Hunter1994}
Furthermore, matches from this group do not show the expected spatial relationship between optical counterparts and maser sources (Figure~\ref{fig:offset_N}).
We would expect masers to be clustered around their parent sources \citep[e.g.,][]{Forster1989,Palagi1993,Walsh2014} with more maser spots closer to the parent source.
Most maser spots should be within 0.5\arcsec of the parent source \citep{Forster1989}.
Matches from this group show no such distribution. 
While matches from this group overlap in Gaia $G$-band apparent magnitude with the young stellar objects identified by \citet{marton2019}, their colours are too blue to be consistent with the YSOs identified by \citet{mowlavi2018}.
We conclude that this colour-magnitude group is dominated by coincidental matches with foreground stars.

The more luminous red quadrant of the colour magnitude diagram we identify as evolved stars.
From studies of open and globular clusters in the Milky Way, the main sequence occupies the lower quadrant between $\rm{G_{abs}} \sim0$, $\rm{G_{bp}-G_{rp}}\lesssim0$ and $\rm{G_{abs}} \lesssim5$, $\rm{G_{bp}-G_{rp}}\lesssim5$.
Stars evolved off the main sequence generally occupy the $\rm{G_{bp}-G_{rp}}\gtrsim1$ and $\rm{G_{abs}} \lesssim4$, occupying the upper right quadrant of the colour magnitude diagram \citep{GAIAHRD2018}.

\begin{figure}
    \centering
    \includegraphics[width=0.45\textwidth]{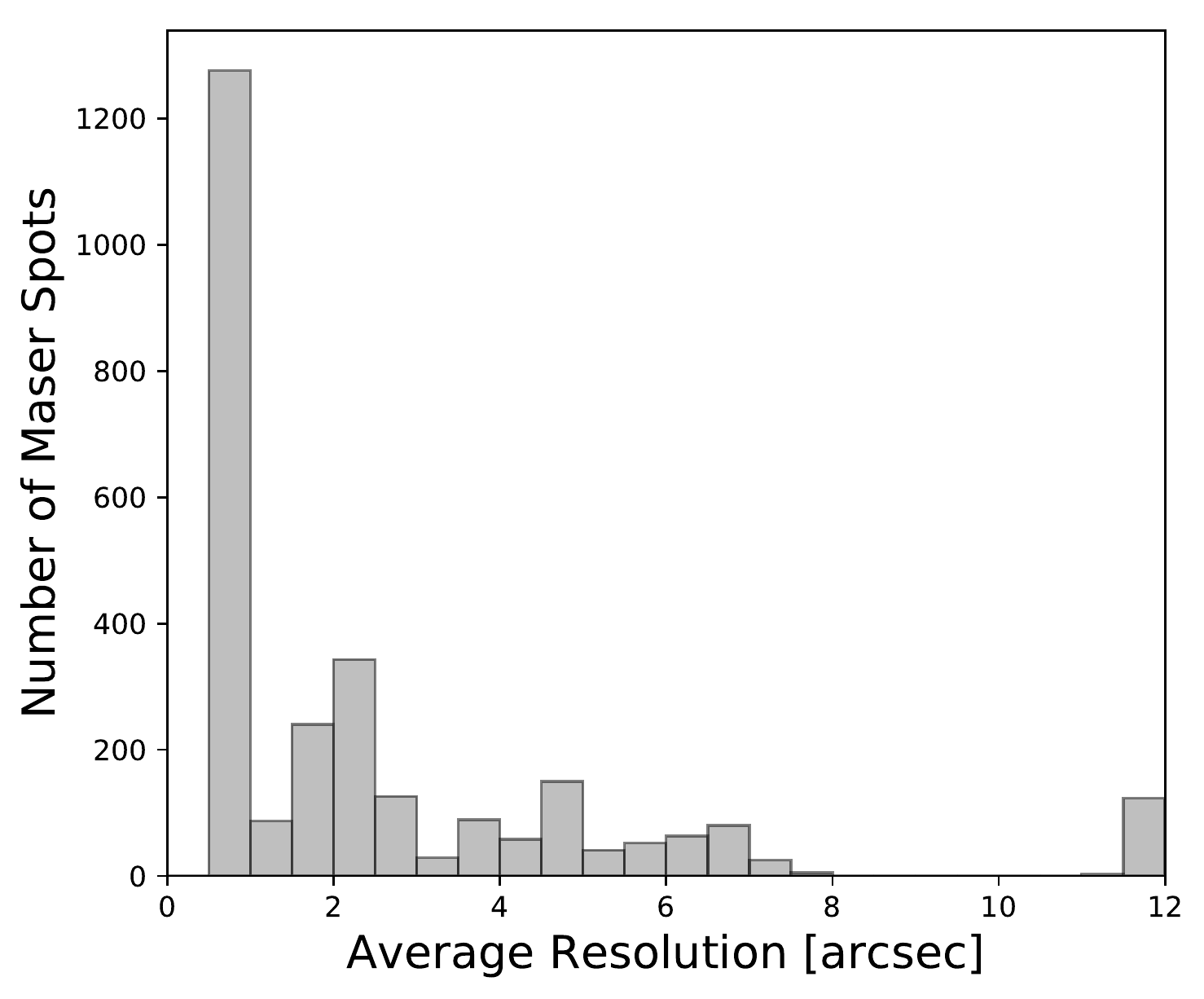}
    \caption{Distribution of spatial resolutions (average between the beam major and minor axes) in the HOPS survey, for sources matched to a Gaia source.
    }
    \label{fig:hopsres}
\end{figure}

\begin{figure}
    \centering
    \includegraphics[width=0.45\textwidth]{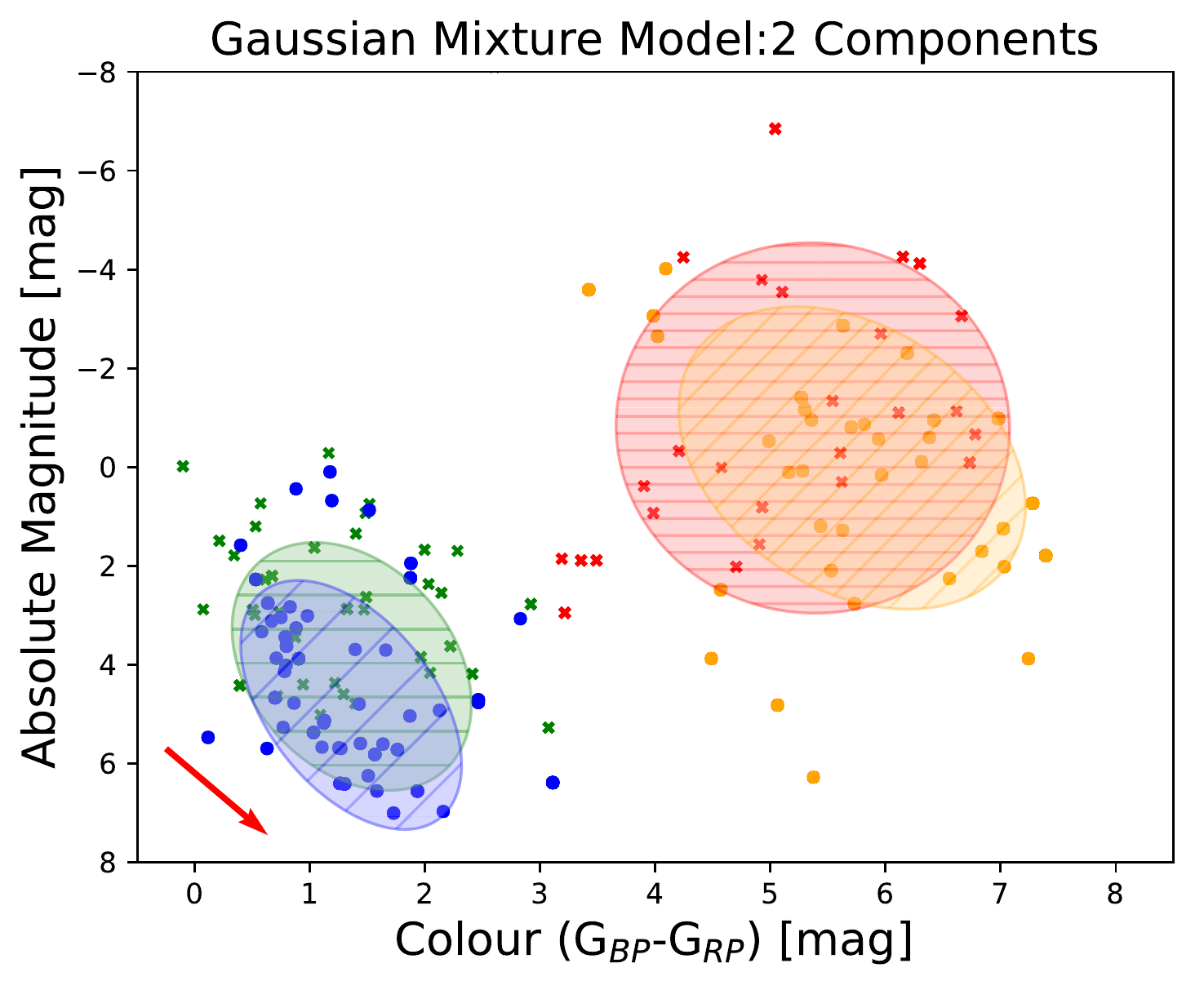}
    \caption{Colour magnitude diagram showing two populations of water masers derived from a two component Gaussian mixture model. This plot includes sources that have not been appropriately extinction corrected, shown with X symbols. 
    Gaussian mixture models including sources not appropriately extinction corrected are shown in red and green with horizontal hatch marks.
    Gaussian mixture models excluding sources not appropriately extinction corrected are shown in orange and blue with diagonal hatch marks.
    Maser sources associated with evolved stars are plotted in orange and red whereas maser sources matched to (but not physically associated with) foreground sources are plotted in blue and green.
    }
    \label{fig:gmm_red_test}
\end{figure}

Masers from evolved stars form in the stellar atmosphere \citep[e.g.,][]{Habing1996,Humphreys2007,Hofner2018}.
Therefore the distribution of  angular offsets between maser site and stellar counterpart would be roughly the convolution of the distributions of resolutions of the Gaia and HOPS catalogues. 
The Full Width Half Maximum (FWHM) of the distribution of the angular offset between Gaia counterpart and the maser should be the convolution of the spatial resolutions of the two surveys. 
Figure~\ref{fig:hopsres} shows the distribution of HOPS spatial resolution for observations of sources with Gaia matches.
45\% of the masers come from observations with a resolution of $<1$\arcsec.
True matches should result in a distribution with a FWHM of 1.12\arcsec.
Our measurement of $1.17\pm0.11$\arcsec (Figure~\ref{fig:offset_N}) is consistent with this expectation. 
The resolution of Gaia is 0.4--0.5\arcsec \citep{GaiaDR22018} and the resolution of HOPS varies between 0.5--15\arcsec, as shown in Figure \ref{fig:hopsres}.
We see from Figure \ref{fig:offset_N} that the distribution of maser spots and stellar counterpart offsets is approximately Gaussian with a FHWM of $\sim$1.
We identify the two groups as the evolved (EV) population and the foreground (FG) population. 
Figure \ref{fig:gmm_red_test} shows the Gaussian mixture model that provides the best fit to the data. 

The extinction vector appears almost parallel to the major axes of the two Gaussian groups.
It may be possible to classify inappropriately extinction corrected sources in the foreground (FG) or evolved (EV) groups. 
To investigate this we first fit a Gaussian mixture model to the entire sample of optical counterparts using the best possible extinction value from \citet{Chen2019}.
The best value means that, for sources with distances beyond the maximum reliable distance found by \citet{Chen2019}, the maximum extinction value is applied to the optical source. 
The results of this fit are shown in Figure \ref{fig:gmm_red_test}.
The original fit is shown with diagonal hatched ellipses.
The horizontal hatched regions show the Gaussian mixture model including the inappropriately extinction corrected sources.
Including the these sources causes the two groups to grow in size, however, no sources changed groups when including all the optical counterparts. We therefore include inappropriately extinction corrected sources in our sample.

\subsubsection{AGB Star Classification} 
\begin{figure}
    \centering
    \includegraphics[width=0.45\textwidth]{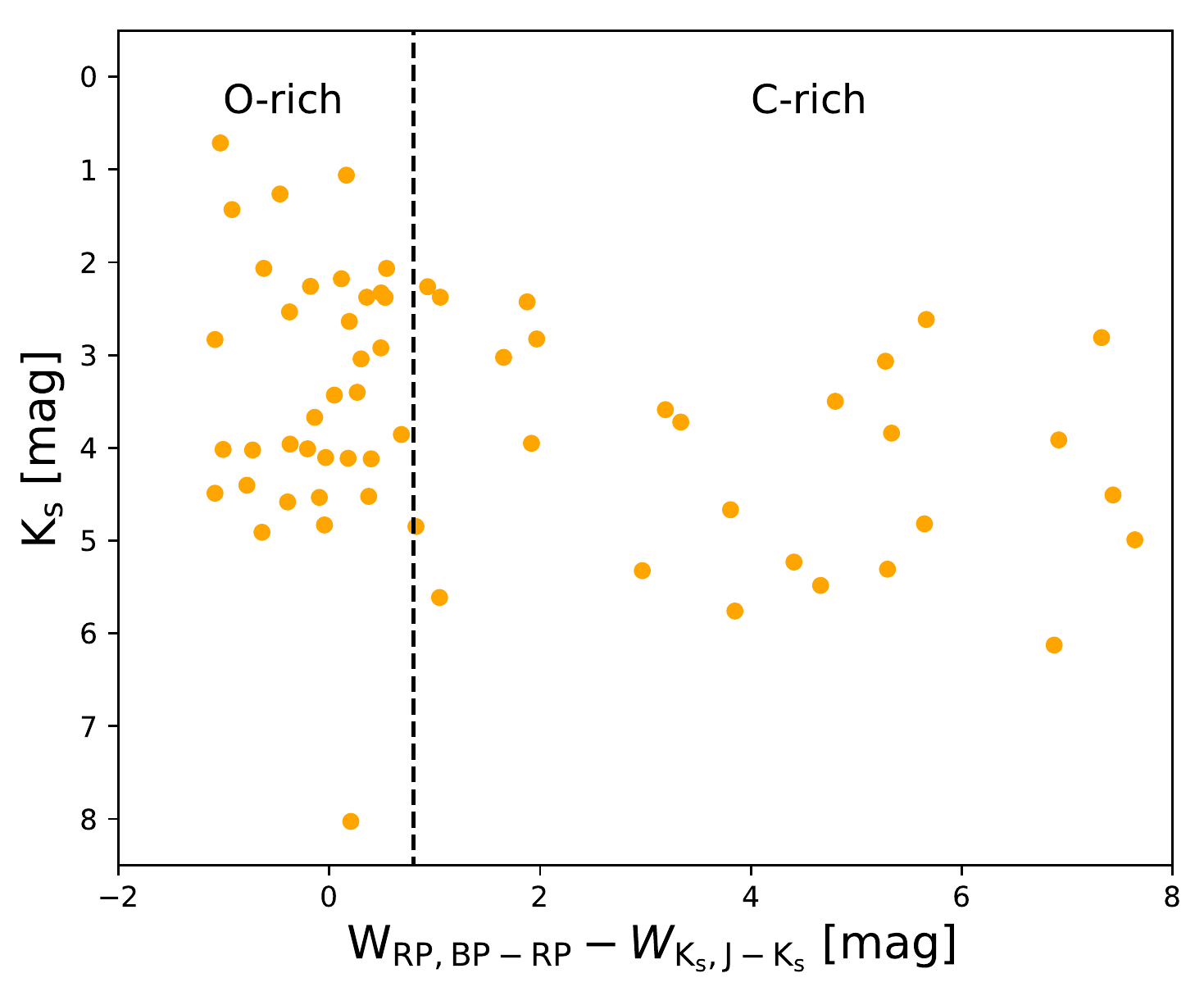}
    \caption{$( \rm{W_{RP,BP-RP}} - \rm{W_{K_s, J-K_s}})$ vs. $\rm{K_s}$ diagram. The vertical dashed line shows the delimeter between o-rich and c-rich AGB stars of 0.8 mag \citep{Lebzelter2018}. } 
    \label{fig:lebzelter}
\end{figure}

It is well established that Asymptotic Giant Branch (AGB) stars host masers in their circumstellar envelopes  (see review by \citealp{Hofner2018} and references therein).
The EV group may be further divided into sub-classes of AGB stars using the multi band approach developed by \citet{Lebzelter2018}.
Using Wesenheit functions from the Gaia BP, RP, and 2MASS J, K$_{\rm s}$ bands, it is possible to identify if the maser stellar counterparts are oxygen rich (O-rich) or carbon rich (C-rich).
The Wesenheit functions are:
\begin{equation}
    \rm{W_{RP,BP-RP}}=\rm{G_{RP}}-1.3\cdot(G_{BP}-G_{RP})
\end{equation}
and
\begin{equation}
    \rm{W_{K_s, J-K_s}}=\rm{K_s - 0.686 \cdot (J-K_s)}
\end{equation}
In the simplest case the C-rich stars have $\rm{W_{RP}}-\rm{W_{K_S}}$ values greater than 0.8 mag.
Figure \ref{fig:lebzelter} shows $( \rm{W_{RP,BP-RP}} - \rm{W_{K_s, J-K_s}})$ vs. $\rm{K_s}$. 
The dashed line shows the delimiter between O-rich and C-rich stellar counterparts.
Of our 62 matched maser sites, 35 are associated with O-rich stars. 
This classification of AGB stars is mostly related to chemistry of the circumstellar envelope, however O-rich stars can have much higher mass loss rates than C-rich stars \citep{Hofner2018}.

\section{Maser Distributions}
\label{maser dist}

\subsection{Galactic Distribution}
\begin{figure}
    \centering
    \includegraphics[width=0.45\textwidth]{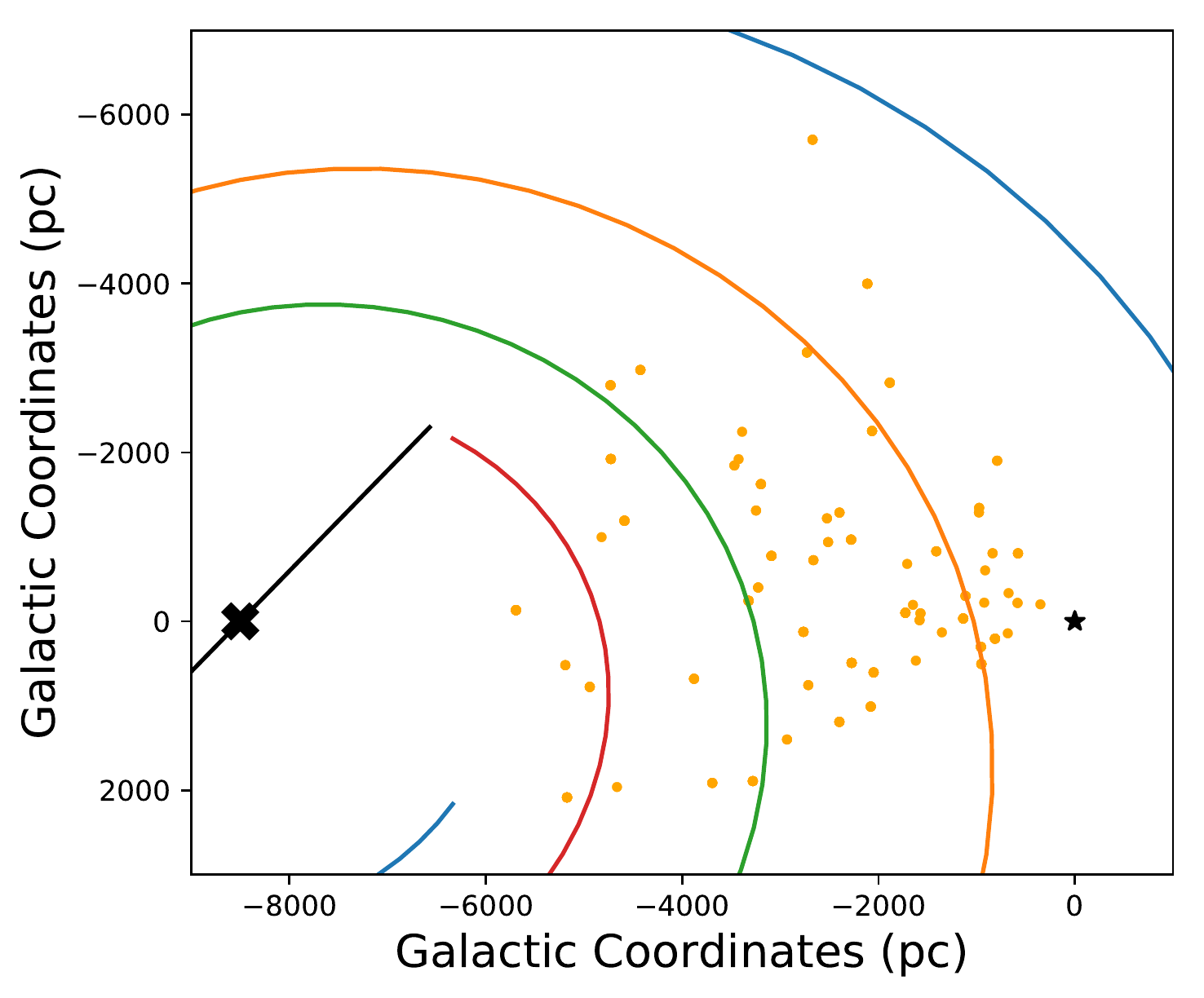}
    \caption{Distribution of the EV maser group in the Milky Way plotted  with orange circles. The red, green, orange, blue, and black lines represent the spiral arms and bar from \citep{vallee2008}. The black X shows the Galactic centre and the black star shows the position of the Solar System.}
    \label{fig:galplot}
\end{figure}

Figure \ref{fig:galplot} shows the distribution of the EV group in the Milky Way. 
This plot includes sources that are not colour and extinction corrected.

\begin{figure}
    \centering
      \includegraphics[width=0.45\textwidth]{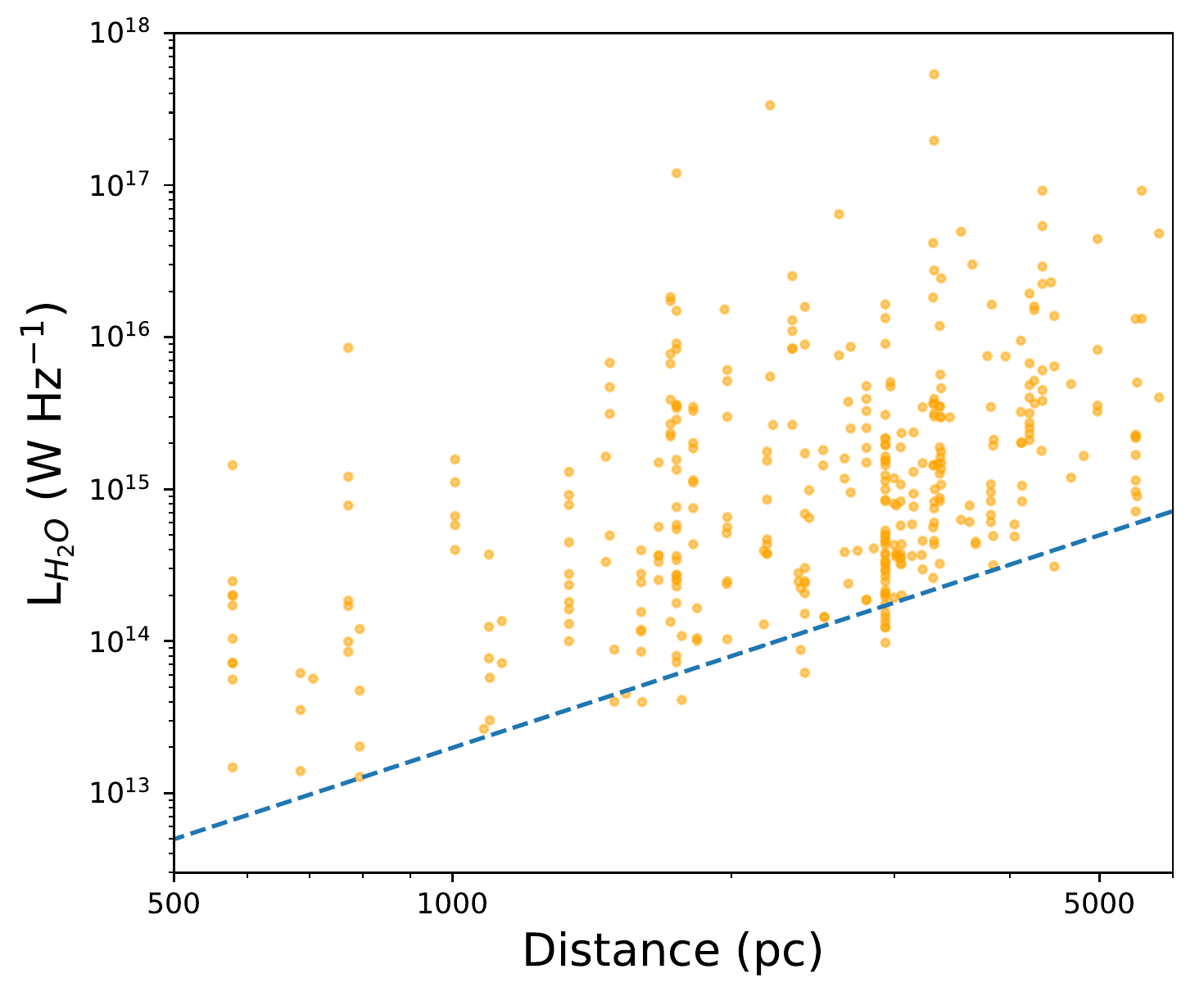}
    \caption{The distance of Gaia-matched maser spots against their peak luminosity. The dotted line represents a flux sensitivity of 0.167~Jy.}
    \label{fig:D_lum_plot}
\end{figure}

In order to estimate the spatial distribution of masers in the Milky Way we need to acknowledge biases in our approach, 
the first being that HOPS is a flux limited survey. Figure \ref{fig:D_lum_plot} shows maser peak luminosity as a function of the distance to the stellar counterpart derived from \citet{BailerJones2018}. 
The dashed line shows the rms noise value of 167~mJy from \citet{Walsh2014}. 
As expected, fewer low-luminosity sources are detected at greater distances. 

A second bias that may be introduced is from the selection of infrared counterparts. Sources that did not have 2MASS counterparts have been removed from our sample. 
This likely happens in the regions with the most extreme extinction (e.g., towards the Galactic centre, \citealp{Dutra2003}). 
Radio frequency masers are not subject to extinction.
Removing the sources with no infrared counterparts thus biases us towards more nearby sources.

\subsubsection{Spiral Arms}
\begin{figure}
    \centering
      \includegraphics[width=0.45\textwidth]{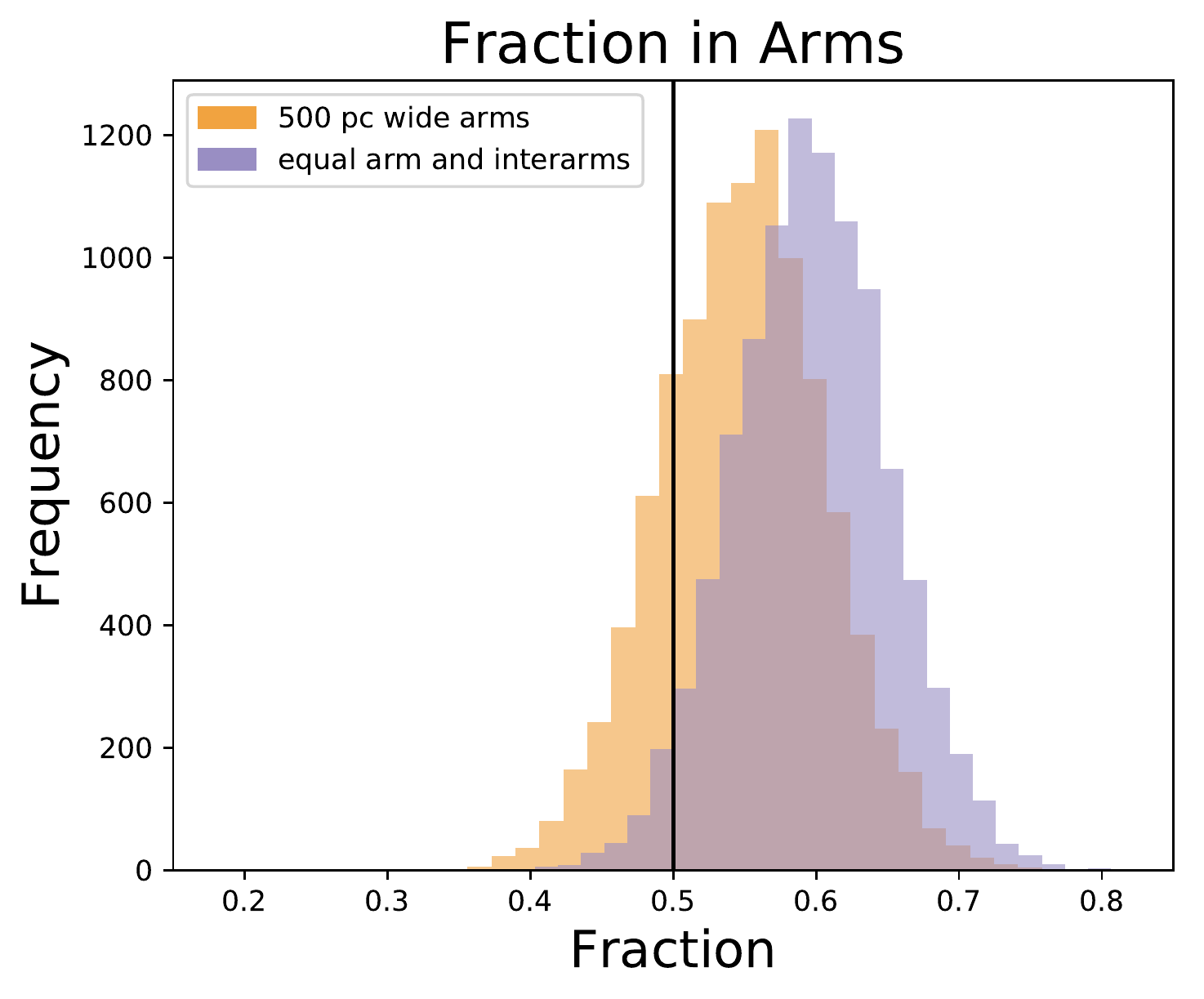}
    \caption{ Fraction of EV masers in spiral arms for 10,000 realisations of EV maser distances determined from Gaia counterparts. The vertical line shows a fraction of 0.5 which would indicate that EV masers are uncorrelated with Galactic structure. In orange we show the distribution using a spiral arm width of 500~pc. In purple we show the distribution of equal arm and inter-arm regions.}
    \label{fig:arm_frac}
\end{figure}

Visual inspection of the EV group maser distribution in the Milky Way does not show a clear spatial correlation with Galactic structure (Fig. \ref{fig:galplot}).
A more quantitative test involves comparing the distribution of masers with a model of Galactic structure.
We use the model of the Milky Way from \citet{vallee2008}, who fit 
a symmetric analytic model to positional and velocity data from dust, stellar clusters, \HI\ and CO emission, finding that the Milky Way is a four armed spiral galaxy,  

Here we use a Monte Carlo method to compute the fraction of EV masers in the spiral arms and inter-arm regions of the Milky Way.
The inter-arms are defined by rotating the symmetric model by 45\deg.
In this manner half of the area of the disk of the Galaxy is defined to be spiral arms and half is defined to be inter-arms.
We calculate the distance to the nearest spiral arm for 10,000 realisations of the distance for each object using the posterior distribution function from \citet{BailerJones2018}. 
A maser is considered to be within a spiral arm if it is closer to the analytic description of the nearest spiral arm instead of the nearest  inter-arm. 
Should the EV masers be uncorrelated with Galactic structure we would expect the fractions of masers in each region (arm or inter-arm) to be 50\%. 
Figure \ref{fig:arm_frac} shows the fraction of masers associated with spiral arms for 10,000 realisations. 
93.3\% of realisations have a fraction of EV masers in the spiral arms greater than 50\%.

The above simple model in which spiral arms comprise of 50\% of the volume of the Galaxy
could be drastically over estimating the number of EV masers associated with spiral arms. 
\citet{Higdon2013} determine the width of the spiral arms to be 500~pc.
They determined this by fitting an axisymmetric disk source model to 205 $\mu$m N{\ts {\scriptsize II}} emission from \hii\ envelopes in the Galactic plane.
The axisymmetric disk source model incorporates a Gaussian parameterization of the arm width.
Performing another 10,000 realizations of the maser distribution in the Milky Way with using a spiral-arm width of 500~pc, we find that 76\% of the realisations have a fraction of EV masers in the spiral arms greater than 50\%. 
These results imply that water masers from evolved stars are concentrated in the spiral arms of the Milky Way.

\section{Discussion}
\label{Discussion}

\subsection{Distribution of Masers}

We have uncovered evidence that \water\ masers associated with evolved stars are associated with Galactic structure.
This is somewhat surprising given that the distribution of AGB stars does not reveal evidence of Galactic structure such as spiral arms \citep{Jackson2002,Robitaille2008}, albeit the structure of the Galaxy may be smoothed out in the analysis of \citet{Jackson2002} due to uncertainty in distance determinations.
Why do evolved stars with masers appear to trace spiral arms when evolved stars without masers do not?

It is possible that the \water\ masers associated with evolved stars are tracing an interaction of the envelope and jets around evolved stars with the molecular gas in spiral arms. 
The lack of  \water\ masers in globular clusters \citep{Cohen1979}, which generally exist outside the disk of the Galaxy and lack an intracluster medium \citep{Barmby2009}, supports this conclusion.
Weaker emission outside the Galactic plane is not unique to \water\ masers.
\citet{Frail1994} find that OH masers in globular clusters are much weaker than those in the disk of the Galaxy. 
The implication is that the galactic environment contributes to the strength of evolved-star \water\ masers.
The strength of the maser is not solely tied to the mass loss rate \citep[e.g.,][]{Engels1986} of the progenitor star.

Mass loss from AGB stars is dominated by a few high mass O-rich objects.
The mass return to the ISM is therefore dominated by O-rich AGB stars \citep{Olofsson2004,Hofner2018}. 
It may be an observational bias that \water\ masers trace the high mass loss objects \citep{Hofner2018}. 
This may explain why we do not observe the smooth distribution of AGB stars from \citet{Jackson2002} and \citet{Robitaille2008}.
Regardless, feedback dominated by the AGB phase of stellar evolution, as traced by water masers, appears to be concentrated in the spiral arms of the Milky Way.

\subsection{Possibilities for improvement}

The initial goal of this work was to determine where \water\ masers are feeding energy into the Galactic environment and the magnitude of said feedback. 
Large unbiased surveys of water masers have yet to cover the entire Galactic plane, with the most complete being  HOPS \citep{Walsh2011} which is limited to 100 square degrees of the inner Galaxy.
If we want to understand how masers impact the Galactic environment as a function of Galactic radius, we would require a complete survey of the Galactic plane.
To determine \water\ maser luminosities and generate a Galactic \water\ maser luminosity function, it is necessary to measure the distances and progenitors to star formation related and evolved star \water\ masers.

We can imagine three ways to generate distance measurement and progenitor properties of \water\ masers.
One would be to perform an unbiased flux limited survey of \water\ maser parallaxes in the Milky Way. 
Maser progenitor information could then be gathered from existing infrared surveys.
This would likely be difficult as stellar \water\ masers are variable on time scales of days \citep[e.g.,][]{Felli2007}.
Another option would be to perform one large flux limited survey of \water\ masers of the Galactic plane without time intensive parallax measurements, and measure parallaxes and stellar counterparts from an infrared Gaia-like telescope with comparable angular resolution; such a telescope does not yet exist.
The third option would be to perform deep surveys of nearby galaxies to determine the distribution and luminosities of their \water\ maser populations. 
This has been attempted \citep[e.g.,][]{Brunthaler2006,Darling2011,Darling2016, Gorski2019}.
However these surveys are limited to only the most luminous masers and are time-intensive.
\citet{Darling2011,Darling2016} spent $\gtrsim1.8$~days of integration on source, mapping 507 pointings in the Andromeda galaxy with the Greenbank Observatory and detected five \water\ masers.
In order to appropriately map the maser distribution in nearby galaxies, new generations of radio telescopes will be needed (e.g., ngVLA; \citealp{McKinnon2019}).

\section{Conclusions}
\label{Conclustions}

\water\ masers are signposts of energy injected into the ISM. 
We have compared the largest available unbiased \water\ maser survey \citep{Walsh2011,Walsh2014} with the second Gaia Data Release \citep{GaiaDR22018}.

Matching \water\ maser sources with optical counterparts yielded a population of masers associated with evolved stars. 
Using a Gaussian mixture model we were able to identify the population of evolved stars despite unsatisfactory extinction corrections. 
We found a slight preference for O-rich AGB stars using the method described by \citet{Lebzelter2018}. 
We could not identify optical counterparts of \water\ masers associated with star formation as these sources are obscured by dust in the Galactic plane.

We compared the distribution of evolved-star \water\ masers to the analytical model of the Milky Way from \citet{vallee2008}. 
We performed a Monte-Carlo simulation of maser positions drawn from the posterior distribution function from \citet{BailerJones2018} with 10$^4$ realisations of the Milky Way's maser distribution.
In the majority of realisations, evolved-star  \water\ masers are positioned closer to spiral arms than to inter-arm regions. 
This suggests a link between masers traced by evolved stars and the spiral pattern of the Milky Way.
This result also suggests that mass loss and feedback from AGB stars is concentrated in the spiral arms of the Milky Way.

\bibliographystyle{mnras}
\bibliography{water_maser.bib,feedback.bib,gaia.bib}

\begin{thebibliography}{}
\makeatletter
\relax
\def\mn@urlcharsother{\let\do\@makeother \do\$\do\&\do\#\do\^\do\_\do\%\do\~}
\def\mn@doi{\begingroup\mn@urlcharsother \@ifnextchar [ {\mn@doi@}
  {\mn@doi@[]}}
\def\mn@doi@[#1]#2{\def\@tempa{#1}\ifx\@tempa\@empty \href
  {http://dx.doi.org/#2} {doi:#2}\else \href {http://dx.doi.org/#2} {#1}\fi
  \endgroup}
\def\mn@eprint#1#2{\mn@eprint@#1:#2::\@nil}
\def\mn@eprint@arXiv#1{\href {http://arxiv.org/abs/#1} {{\tt arXiv:#1}}}
\def\mn@eprint@dblp#1{\href {http://dblp.uni-trier.de/rec/bibtex/#1.xml}
  {dblp:#1}}
\def\mn@eprint@#1:#2:#3:#4\@nil{\def\@tempa {#1}\def\@tempb {#2}\def\@tempc
  {#3}\ifx \@tempc \@empty \let \@tempc \@tempb \let \@tempb \@tempa \fi \ifx
  \@tempb \@empty \def\@tempb {arXiv}\fi \@ifundefined
  {mn@eprint@\@tempb}{\@tempb:\@tempc}{\expandafter \expandafter \csname
  mn@eprint@\@tempb\endcsname \expandafter{\@tempc}}}

\bibitem[\protect\citeauthoryear{{Arenou} et~al.,}{{Arenou}
  et~al.}{2018}]{Arenou2018}
{Arenou} F.,  et~al., 2018, \mn@doi [\aap] {10.1051/0004-6361/201833234}, \href
  {https://ui.adsabs.harvard.edu/abs/2018A&A...616A..17A} {616, A17}

\bibitem[\protect\citeauthoryear{{Bailer-Jones}, {Rybizki}, {Fouesneau},
  {Mantelet}  \& {Andrae}}{{Bailer-Jones} et~al.}{2018}]{BailerJones2018}
{Bailer-Jones} C.~A.~L.,  {Rybizki} J.,  {Fouesneau} M.,  {Mantelet} G.,
  {Andrae} R.,  2018, \mn@doi [\aj] {10.3847/1538-3881/aacb21}, \href
  {http://adsabs.harvard.edu/abs/2018AJ....156...58B} {156, 58}

\bibitem[\protect\citeauthoryear{{Barmby}, {Boyer}, {Woodward}, {Gehrz}, {van
  Loon}, {Fazio}, {Marengo}  \& {Polomski}}{{Barmby} et~al.}{2009}]{Barmby2009}
{Barmby} P.,  {Boyer} M.~L.,  {Woodward} C.~E.,  {Gehrz} R.~D.,  {van Loon}
  J.~T.,  {Fazio} G.~G.,  {Marengo} M.,   {Polomski} E.,  2009, \mn@doi [\aj]
  {10.1088/0004-6256/137/1/207}, 137, 207

\bibitem[\protect\citeauthoryear{{Brand}}{{Brand}}{1986}]{Brand1986}
{Brand} J.,  1986, PhD thesis, Leiden Univ., Netherlands.

\bibitem[\protect\citeauthoryear{{Brunthaler}, {Henkel}, {de Blok}, {Reid},
  {Greenhill}  \& {Falcke}}{{Brunthaler} et~al.}{2006}]{Brunthaler2006}
{Brunthaler} A.,  {Henkel} C.,  {de Blok} W.~J.~G.,  {Reid} M.~J.,  {Greenhill}
  L.~J.,   {Falcke} H.,  2006, \mn@doi [\aap] {10.1051/0004-6361:20065650},
  \href {https://ui.adsabs.harvard.edu/abs/2006A&A...457..109B} {457, 109}

\bibitem[\protect\citeauthoryear{{Cesaroni}, {Palagi}, {Felli}, {Catarzi},
  {Comoretto}, {Di Franco}, {Giovanardi}  \& {Palla}}{{Cesaroni}
  et~al.}{1988}]{Cesaroni1988}
{Cesaroni} R.,  {Palagi} F.,  {Felli} M.,  {Catarzi} M.,  {Comoretto} G.,  {Di
  Franco} S.,  {Giovanardi} C.,   {Palla} F.,  1988, \aaps, \href
  {https://ui.adsabs.harvard.edu/abs/1988A&AS...76..445C} {76, 445}

\bibitem[\protect\citeauthoryear{{Chen} et~al.,}{{Chen}
  et~al.}{2017}]{Chen2017}
{Chen} B.-Q.,  et~al., 2017, \mn@doi [\mnras] {10.1093/mnras/stw2497}, \href
  {http://adsabs.harvard.edu/abs/2017MNRAS.464.2545C} {464, 2545}

\bibitem[\protect\citeauthoryear{{Chen} et~al.,}{{Chen}
  et~al.}{2019}]{Chen2019}
{Chen} B.-Q.,  et~al., 2019, \mn@doi [\mnras] {10.1093/mnras/sty3341}, \href
  {http://adsabs.harvard.edu/abs/2019MNRAS.483.4277C} {483, 4277}

\bibitem[\protect\citeauthoryear{{Cohen} \& {Malkan}}{{Cohen} \&
  {Malkan}}{1979}]{Cohen1979}
{Cohen} N.~L.,  {Malkan} M.~A.,  1979, \mn@doi [\aj] {10.1086/112390}, \href
  {https://ui.adsabs.harvard.edu/abs/1979AJ.....84...74C} {84, 74}

\bibitem[\protect\citeauthoryear{{Comoretto} et~al.,}{{Comoretto}
  et~al.}{1990}]{Comoretto1990}
{Comoretto} G.,  et~al., 1990, \aaps, \href
  {http://adsabs.harvard.edu/abs/1990A%26AS...84..179C} {84, 179}

\bibitem[\protect\citeauthoryear{{Darling}}{{Darling}}{2011}]{Darling2011}
{Darling} J.,  2011, \mn@doi [\apjl] {10.1088/2041-8205/732/1/L2}, \href
  {https://ui.adsabs.harvard.edu/abs/2011ApJ...732L...2D} {732, L2}

\bibitem[\protect\citeauthoryear{{Darling}, {Gerard}, {Amiri}  \&
  {Lawrence}}{{Darling} et~al.}{2016}]{Darling2016}
{Darling} J.,  {Gerard} B.,  {Amiri} N.,   {Lawrence} K.,  2016, \mn@doi [\apj]
  {10.3847/0004-637X/826/1/24}, \href
  {https://ui.adsabs.harvard.edu/abs/2016ApJ...826...24D} {826, 24}

\bibitem[\protect\citeauthoryear{{Dickinson}, {Kojoian}  \&
  {Strom}}{{Dickinson} et~al.}{1974}]{Dickinson1974}
{Dickinson} D.~F.,  {Kojoian} G.,   {Strom} S.~E.,  1974, \mn@doi [\apjl]
  {10.1086/181677}, \href {http://adsabs.harvard.edu/abs/1974ApJ...194L..93D}
  {194, L93}

\bibitem[\protect\citeauthoryear{{Dutra}, {Santiago}, {Bica}  \&
  {Barbuy}}{{Dutra} et~al.}{2003}]{Dutra2003}
{Dutra} C.~M.,  {Santiago} B.~X.,  {Bica} E.~L.~D.,   {Barbuy} B.,  2003,
  \mn@doi [\mnras] {10.1046/j.1365-8711.2003.06049.x}, \href
  {https://ui.adsabs.harvard.edu/abs/2003MNRAS.338..253D} {338, 253}

\bibitem[\protect\citeauthoryear{{Elitzur}, {Hollenbach}  \& {McKee}}{{Elitzur}
  et~al.}{1989}]{Elitzur1989}
{Elitzur} M.,  {Hollenbach} D.~J.,   {McKee} C.~F.,  1989, \mn@doi [\apj]
  {10.1086/168080}, \href {http://adsabs.harvard.edu/abs/1989ApJ...346..983E}
  {346, 983}

\bibitem[\protect\citeauthoryear{{Engels}, {Schmid-Burgk}  \&
  {Walmsley}}{{Engels} et~al.}{1986}]{Engels1986}
{Engels} D.,  {Schmid-Burgk} J.,   {Walmsley} C.~M.,  1986, \aap, \href
  {http://adsabs.harvard.edu/abs/1986A%26A...167..129E} {167, 129}

\bibitem[\protect\citeauthoryear{{Felli} et~al.,}{{Felli}
  et~al.}{2007}]{Felli2007}
{Felli} M.,  et~al., 2007, \mn@doi [\aap] {10.1051/0004-6361:20077804}, \href
  {https://ui.adsabs.harvard.edu/abs/2007A&A...476..373F} {476, 373}

\bibitem[\protect\citeauthoryear{{Forster} \& {Caswell}}{{Forster} \&
  {Caswell}}{1989}]{Forster1989}
{Forster} J.~R.,  {Caswell} J.~L.,  1989, \aap, \href
  {http://adsabs.harvard.edu/abs/1989A%26A...213..339F} {213, 339}

\bibitem[\protect\citeauthoryear{{Frail} \& {Beasley}}{{Frail} \&
  {Beasley}}{1994}]{Frail1994}
{Frail} D.~A.,  {Beasley} A.~J.,  1994, \aap, \href
  {https://ui.adsabs.harvard.edu/abs/1994A&A...290..796F} {290, 796}

\bibitem[\protect\citeauthoryear{{Furuya}, {Kitamura}, {Wootten}, {Claussen}
  \& {Kawabe}}{{Furuya} et~al.}{2003}]{Furuya2003}
{Furuya} R.~S.,  {Kitamura} Y.,  {Wootten} A.,  {Claussen} M.~J.,   {Kawabe}
  R.,  2003, \mn@doi [\apjs] {10.1086/342749}, \href
  {http://adsabs.harvard.edu/abs/2003ApJS..144...71F} {144, 71}

\bibitem[\protect\citeauthoryear{{Furuya}, {Kitamura}, {Wootten}, {Claussen}
  \& {Kawabe}}{{Furuya} et~al.}{2007a}]{Furuya2007a}
{Furuya} R.~S.,  {Kitamura} Y.,  {Wootten} A.,  {Claussen} M.~J.,   {Kawabe}
  R.,  2007a, \mn@doi [\apjs] {10.1086/516825}, \href
  {http://adsabs.harvard.edu/abs/2007ApJS..171..349F} {171, 349}

\bibitem[\protect\citeauthoryear{{Furuya}, {Kitamura}, {Wootten}, {Claussen}
  \& {Kawabe}}{{Furuya} et~al.}{2007b}]{Furuya2007b}
{Furuya} R.~S.,  {Kitamura} Y.,  {Wootten} H.~A.,  {Claussen} M.~J.,   {Kawabe}
  R.,  2007b, \mn@doi [\apjl] {10.1086/516822}, \href
  {http://adsabs.harvard.edu/abs/2007ApJ...659L..81F} {659, L81}

\bibitem[\protect\citeauthoryear{{Gaia Collaboration} et~al.,}{{Gaia
  Collaboration} et~al.}{2018a}]{GaiaDR22018}
{Gaia Collaboration} et~al., 2018a, \mn@doi [\aap]
  {10.1051/0004-6361/201833051}, \href
  {http://adsabs.harvard.edu/abs/2018A%26A...616A...1G} {616, A1}

\bibitem[\protect\citeauthoryear{{Gaia Collaboration} et~al.,}{{Gaia
  Collaboration} et~al.}{2018b}]{GAIAHRD2018}
{Gaia Collaboration} et~al., 2018b, \mn@doi [\aap]
  {10.1051/0004-6361/201832843}, \href
  {https://ui.adsabs.harvard.edu/abs/2018A&A...616A..10G} {616, A10}

\bibitem[\protect\citeauthoryear{{Genzel} \& {Downes}}{{Genzel} \&
  {Downes}}{1977}]{Genzel1977}
{Genzel} R.,  {Downes} D.,  1977, \aaps, \href
  {http://adsabs.harvard.edu/abs/1977A%26AS...30..145G} {30, 145}

\bibitem[\protect\citeauthoryear{{Gorski}, {Ott}, {Rand}, {Meier}, {Momjian}
  \& {Schinnerer}}{{Gorski} et~al.}{2017}]{Gorski2017}
{Gorski} M.,  {Ott} J.,  {Rand} R.,  {Meier} D.~S.,  {Momjian} E.,
  {Schinnerer} E.,  2017, \mn@doi [\apj] {10.3847/1538-4357/aa74af}, \href
  {http://adsabs.harvard.edu/abs/2017ApJ...842..124G} {842, 124}

\bibitem[\protect\citeauthoryear{{Gorski}, {Ott}, {Rand}, {Meier}, {Momjian}
  \& {Schinnerer}}{{Gorski} et~al.}{2018}]{Gorski2018}
{Gorski} M.,  {Ott} J.,  {Rand} R.,  {Meier} D.~S.,  {Momjian} E.,
  {Schinnerer} E.,  2018, \mn@doi [\apj] {10.3847/1538-4357/aab3cc}, \href
  {http://adsabs.harvard.edu/abs/2018ApJ...856..134G} {856, 134}

\bibitem[\protect\citeauthoryear{{Gorski}, {Ott}, {Rand}, {Meier}, {Momjian},
  {Schinnerer}  \& {Ellingsen}}{{Gorski} et~al.}{2019}]{Gorski2019}
{Gorski} M.~D.,  {Ott} J.,  {Rand} R.,  {Meier} D.~S.,  {Momjian} E.,
  {Schinnerer} E.,   {Ellingsen} S.~P.,  2019, \mn@doi [\mnras]
  {10.1093/mnras/sty3077}, \href
  {http://adsabs.harvard.edu/abs/2019MNRAS.483.5434G} {483, 5434}

\bibitem[\protect\citeauthoryear{{Habing}}{{Habing}}{1996}]{Habing1996}
{Habing} H.~J.,  1996, \mn@doi [\aapr] {10.1007/PL00013287}, \href
  {https://ui.adsabs.harvard.edu/abs/1996A&ARv...7...97H} {7, 97}

\bibitem[\protect\citeauthoryear{{Hagiwara}, {Henkel}, {Menten}  \&
  {Nakai}}{{Hagiwara} et~al.}{2001}]{Hagiwara2001}
{Hagiwara} Y.,  {Henkel} C.,  {Menten} K.~M.,   {Nakai} N.,  2001, \mn@doi
  [\apjl] {10.1086/324171}, \href
  {http://adsabs.harvard.edu/abs/2001ApJ...560L..37H} {560, L37}

\bibitem[\protect\citeauthoryear{{Herrnstein} et~al.,}{{Herrnstein}
  et~al.}{1999}]{Herrnstien1999}
{Herrnstein} J.~R.,  et~al., 1999, \mn@doi [\nat] {10.1038/22972}, \href
  {http://adsabs.harvard.edu/abs/1999Natur.400..539H} {400, 539}

\bibitem[\protect\citeauthoryear{{Higdon} \& {Lingenfelter}}{{Higdon} \&
  {Lingenfelter}}{2013}]{Higdon2013}
{Higdon} J.~C.,  {Lingenfelter} R.~E.,  2013, \mn@doi [\apj]
  {10.1088/0004-637X/775/2/110}, \href
  {https://ui.adsabs.harvard.edu/abs/2013ApJ...775..110H} {775, 110}

\bibitem[\protect\citeauthoryear{{H{\"o}fner} \& {Olofsson}}{{H{\"o}fner} \&
  {Olofsson}}{2018}]{Hofner2018}
{H{\"o}fner} S.,  {Olofsson} H.,  2018, \mn@doi [\aapr]
  {10.1007/s00159-017-0106-5}, \href
  {https://ui.adsabs.harvard.edu/abs/2018A&ARv..26....1H} {26, 1}

\bibitem[\protect\citeauthoryear{{Hopkins}, {Quataert}  \& {Murray}}{{Hopkins}
  et~al.}{2011}]{Hopkins2011}
{Hopkins} P.~F.,  {Quataert} E.,   {Murray} N.,  2011, \mn@doi [\mnras]
  {10.1111/j.1365-2966.2011.19306.x}, \href
  {http://adsabs.harvard.edu/abs/2011MNRAS.417..950H} {417, 950}

\bibitem[\protect\citeauthoryear{{Hopkins}, {Quataert}  \& {Murray}}{{Hopkins}
  et~al.}{2012}]{Hopkins2012}
{Hopkins} P.~F.,  {Quataert} E.,   {Murray} N.,  2012, \mn@doi [\mnras]
  {10.1111/j.1365-2966.2012.20578.x}, \href
  {http://adsabs.harvard.edu/abs/2012MNRAS.421.3488H} {421, 3488}

\bibitem[\protect\citeauthoryear{{Hopkins}, {Kere{\v s}}, {O{\~n}orbe},
  {Faucher-Gigu{\`e}re}, {Quataert}, {Murray}  \& {Bullock}}{{Hopkins}
  et~al.}{2014}]{Hopkins2014}
{Hopkins} P.~F.,  {Kere{\v s}} D.,  {O{\~n}orbe} J.,  {Faucher-Gigu{\`e}re}
  C.-A.,  {Quataert} E.,  {Murray} N.,   {Bullock} J.~S.,  2014, \mn@doi
  [\mnras] {10.1093/mnras/stu1738}, \href
  {http://adsabs.harvard.edu/abs/2014MNRAS.445..581H} {445, 581}

\bibitem[\protect\citeauthoryear{{Humphreys}}{{Humphreys}}{2007}]{Humphreys2007}
{Humphreys} E.~M.~L.,  2007, in {Chapman} J.~M.,  {Baan} W.~A.,  eds,  IAU
  Symposium Vol. 242, Astrophysical Masers and their Environments. pp 471--480
  (\mn@eprint {arXiv} {0705.4456}), \mn@doi{10.1017/S1743921307013622}

\bibitem[\protect\citeauthoryear{{Hunter}, {Taylor}, {Felli}  \&
  {Tofani}}{{Hunter} et~al.}{1994}]{Hunter1994}
{Hunter} T.~R.,  {Taylor} G.~B.,  {Felli} M.,   {Tofani} G.,  1994, \aap, \href
  {https://ui.adsabs.harvard.edu/abs/1994A&A...284..215H} {284, 215}

\bibitem[\protect\citeauthoryear{{Jackson}, {Ivezi{\'c}}  \& {Knapp}}{{Jackson}
  et~al.}{2002}]{Jackson2002}
{Jackson} T.,  {Ivezi{\'c}} {\v{Z}}.,   {Knapp} G.~R.,  2002, \mn@doi [\mnras]
  {10.1046/j.1365-8711.2002.05980.x}, \href
  {https://ui.adsabs.harvard.edu/abs/2002MNRAS.337..749J} {337, 749}

\bibitem[\protect\citeauthoryear{{Kauffmann}, {Colberg}, {Diaferio}  \&
  {White}}{{Kauffmann} et~al.}{1999}]{Kauffmann1999}
{Kauffmann} G.,  {Colberg} J.~M.,  {Diaferio} A.,   {White} S.~D.~M.,  1999,
  \mn@doi [\mnras] {10.1046/j.1365-8711.1999.02202.x}, \href
  {http://adsabs.harvard.edu/abs/1999MNRAS.303..188K} {303, 188}

\bibitem[\protect\citeauthoryear{{Kennicutt} Jr. et~al.,}{{Kennicutt}
  et~al.}{2007}]{Kennicutt2007}
{Kennicutt} Jr. R.~C.,  et~al., 2007, \mn@doi [\apj] {10.1086/522300}, \href
  {http://adsabs.harvard.edu/abs/2007ApJ...671..333K} {671, 333}

\bibitem[\protect\citeauthoryear{{Krumholz}, {Klein}  \& {McKee}}{{Krumholz}
  et~al.}{2011}]{Krumholz2011}
{Krumholz} M.~R.,  {Klein} R.~I.,   {McKee} C.~F.,  2011, \mn@doi [\apj]
  {10.1088/0004-637X/740/2/74}, \href
  {http://adsabs.harvard.edu/abs/2011ApJ...740...74K} {740, 74}

\bibitem[\protect\citeauthoryear{{Lebzelter}, {Mowlavi}, {Marigo},
  {Pastorelli}, {Trabucchi}, {Wood}  \& {Lecoeur-Ta{\"\i}bi}}{{Lebzelter}
  et~al.}{2018}]{Lebzelter2018}
{Lebzelter} T.,  {Mowlavi} N.,  {Marigo} P.,  {Pastorelli} G.,  {Trabucchi} M.,
   {Wood} P.~R.,   {Lecoeur-Ta{\"\i}bi} I.,  2018, \mn@doi [\aap]
  {10.1051/0004-6361/201833615}, \href
  {https://ui.adsabs.harvard.edu/abs/2018A&A...616L..13L} {616, L13}

\bibitem[\protect\citeauthoryear{{Leroy} et~al.,}{{Leroy}
  et~al.}{2013}]{Leroy2013}
{Leroy} A.~K.,  et~al., 2013, \mn@doi [\aj] {10.1088/0004-6256/146/2/19}, \href
  {http://adsabs.harvard.edu/abs/2013AJ....146...19L} {146, 19}

\bibitem[\protect\citeauthoryear{{Marrese}, {Marinoni}, {Fabrizio}  \&
  {Giuffrida}}{{Marrese} et~al.}{2017}]{Marrese2017}
{Marrese} P.~M.,  {Marinoni} S.,  {Fabrizio} M.,   {Giuffrida} G.,  2017,
  \mn@doi [\aap] {10.1051/0004-6361/201730965}, \href
  {http://adsabs.harvard.edu/abs/2017A%26A...607A.105M} {607, A105}

\bibitem[\protect\citeauthoryear{{Marrese}, {Marinoni}, {Fabrizio}  \&
  {Altavilla}}{{Marrese} et~al.}{2019}]{Marrese2019}
{Marrese} P.~M.,  {Marinoni} S.,  {Fabrizio} M.,   {Altavilla} G.,  2019,
  \mn@doi [\aap] {10.1051/0004-6361/201834142}, \href
  {http://adsabs.harvard.edu/abs/2019A%26A...621A.144M} {621, A144}

\bibitem[\protect\citeauthoryear{{Marton} et~al.,}{{Marton}
  et~al.}{2019}]{marton2019}
{Marton} G.,  et~al., 2019, \mn@doi [\mnras] {10.1093/mnras/stz1301}, \href
  {https://ui.adsabs.harvard.edu/abs/2019MNRAS.487.2522M} {487, 2522}

\bibitem[\protect\citeauthoryear{{McKinnon}, {Beasley}, {Murphy}, {Selina},
  {Farnsworth}  \& {Walter}}{{McKinnon} et~al.}{2019}]{McKinnon2019}
{McKinnon} M.,  {Beasley} A.,  {Murphy} E.,  {Selina} R.,  {Farnsworth} R.,
  {Walter} A.,  2019, in \baas. p.~81

\bibitem[\protect\citeauthoryear{{Meidt} et~al.,}{{Meidt}
  et~al.}{2015}]{Meidt2015}
{Meidt} S.~E.,  et~al., 2015, \mn@doi [\apj] {10.1088/0004-637X/806/1/72},
  \href {http://adsabs.harvard.edu/abs/2015ApJ...806...72M} {806, 72}

\bibitem[\protect\citeauthoryear{{Miranda}, {G{\'o}mez}, {Anglada}  \&
  {Torrelles}}{{Miranda} et~al.}{2001}]{Miranda2001}
{Miranda} L.~F.,  {G{\'o}mez} Y.,  {Anglada} G.,   {Torrelles} J.~M.,  2001,
  \mn@doi [\nat] {10.1038/35104518}, \href
  {http://adsabs.harvard.edu/abs/2001Natur.414..284M} {414, 284}

\bibitem[\protect\citeauthoryear{{Moore}, {Mountain}, {Yamashita}  \&
  {Selby}}{{Moore} et~al.}{1988}]{Moore1988}
{Moore} T.~J.~T.,  {Mountain} C.~M.,  {Yamashita} T.,   {Selby} M.~J.,  1988,
  \mn@doi [\mnras] {10.1093/mnras/234.1.95}, \href
  {https://ui.adsabs.harvard.edu/abs/1988MNRAS.234...95M} {234, 95}

\bibitem[\protect\citeauthoryear{{Mowlavi} et~al.,}{{Mowlavi}
  et~al.}{2018}]{mowlavi2018}
{Mowlavi} N.,  et~al., 2018, \mn@doi [\aap] {10.1051/0004-6361/201833366},
  \href {https://ui.adsabs.harvard.edu/abs/2018A&A...618A..58M} {618, A58}

\bibitem[\protect\citeauthoryear{Olofsson}{Olofsson}{2004}]{Olofsson2004}
Olofsson H.,  2004, Circumstellar Envelopes.
Springer New York, New York, NY, pp 325--410,
  \mn@doi{10.1007/978-1-4757-3876-6_7}

\bibitem[\protect\citeauthoryear{Palagi, Cesaroni, Comoretto, Felli  \&
  Natale}{Palagi et~al.}{1993}]{Palagi1993}
Palagi F.,  Cesaroni R.,  Comoretto G.,  Felli M.,   Natale V.,  1993,
  Astronomy and Astrophysics Supplement, 101, 153

\bibitem[\protect\citeauthoryear{Pedregosa et~al.,}{Pedregosa
  et~al.}{2011}]{scikit-learn}
Pedregosa F.,  et~al., 2011, Journal of Machine Learning Research, 12, 2825

\bibitem[\protect\citeauthoryear{{Pihlstr{\"o}m}, {Sjouwerman}, {Claussen},
  {Morris}, {Rich}, {van Langevelde}  \& {Quiroga-Nu{\~n}ez}}{{Pihlstr{\"o}m}
  et~al.}{2018}]{Pihlstrom2018}
{Pihlstr{\"o}m} Y.~M.,  {Sjouwerman} L.~O.,  {Claussen} M.~J.,  {Morris} M.~R.,
   {Rich} R.~M.,  {van Langevelde} H.~J.,   {Quiroga-Nu{\~n}ez} L.~H.,  2018,
  \mn@doi [\apj] {10.3847/1538-4357/aae77d}, \href
  {https://ui.adsabs.harvard.edu/abs/2018ApJ...868...72P} {868, 72}

\bibitem[\protect\citeauthoryear{{Reid}, {Braatz}, {Condon}, {Greenhill},
  {Henkel}  \& {Lo}}{{Reid} et~al.}{2009}]{Reid2009}
{Reid} M.~J.,  {Braatz} J.~A.,  {Condon} J.~J.,  {Greenhill} L.~J.,  {Henkel}
  C.,   {Lo} K.~Y.,  2009, \mn@doi [\apj] {10.1088/0004-637X/695/1/287}, \href
  {http://adsabs.harvard.edu/abs/2009ApJ...695..287R} {695, 287}

\bibitem[\protect\citeauthoryear{{Robitaille} et~al.,}{{Robitaille}
  et~al.}{2008}]{Robitaille2008}
{Robitaille} T.~P.,  et~al., 2008, \mn@doi [\aj]
  {10.1088/0004-6256/136/6/2413}, \href
  {https://ui.adsabs.harvard.edu/abs/2008AJ....136.2413R} {136, 2413}

\bibitem[\protect\citeauthoryear{{Skrutskie} et~al.,}{{Skrutskie}
  et~al.}{2006}]{Skrutskie2006}
{Skrutskie} M.~F.,  et~al., 2006, \mn@doi [\aj] {10.1086/498708}, \href
  {http://adsabs.harvard.edu/abs/2006AJ....131.1163S} {131, 1163}

\bibitem[\protect\citeauthoryear{{Taylor}}{{Taylor}}{2005}]{Taylor2005}
{Taylor} M.~B.,  2005, in {Shopbell} P.,  {Britton} M.,   {Ebert} R.,  eds,
  Astronomical Society of the Pacific Conference Series Vol. 347, Astronomical
  Data Analysis Software and Systems XIV. p.~29

\bibitem[\protect\citeauthoryear{{Titmarsh}, {Ellingsen}, {Breen}, {Caswell}
  \& {Voronkov}}{{Titmarsh} et~al.}{2013}]{Titmarsh2013}
{Titmarsh} A.~M.,  {Ellingsen} S.~P.,  {Breen} S.~L.,  {Caswell} J.~L.,
  {Voronkov} M.~A.,  2013, \mn@doi [\apjl] {10.1088/2041-8205/775/1/L12}, \href
  {http://adsabs.harvard.edu/abs/2013ApJ...775L..12T} {775, L12}

\bibitem[\protect\citeauthoryear{{Urquhart} et~al.,}{{Urquhart}
  et~al.}{2009}]{Urquhart2009}
{Urquhart} J.~S.,  et~al., 2009, \mn@doi [\aap] {10.1051/0004-6361/200912608},
  \href {http://adsabs.harvard.edu/abs/2009A%26A...507..795U} {507, 795}

\bibitem[\protect\citeauthoryear{{Urquhart} et~al.,}{{Urquhart}
  et~al.}{2011}]{Urquhart2011}
{Urquhart} J.~S.,  et~al., 2011, \mn@doi [\mnras]
  {10.1111/j.1365-2966.2011.19594.x}, \href
  {http://adsabs.harvard.edu/abs/2011MNRAS.418.1689U} {418, 1689}

\bibitem[\protect\citeauthoryear{{Vall{\'e}e}}{{Vall{\'e}e}}{2008}]{vallee2008}
{Vall{\'e}e} J.~P.,  2008, \mn@doi [\aj] {10.1088/0004-6256/135/4/1301}, \href
  {http://adsabs.harvard.edu/abs/2008AJ....135.1301V} {135, 1301}

\bibitem[\protect\citeauthoryear{{Walsh} et~al.,}{{Walsh}
  et~al.}{2011}]{Walsh2011}
{Walsh} A.~J.,  et~al., 2011, \mn@doi [\mnras]
  {10.1111/j.1365-2966.2011.19115.x}, \href
  {https://ui.adsabs.harvard.edu/abs/2011MNRAS.416.1764W} {416, 1764}

\bibitem[\protect\citeauthoryear{{Walsh}, {Purcell}, {Longmore}, {Breen},
  {Green}, {Harvey-Smith}, {Jordan}  \& {Macpherson}}{{Walsh}
  et~al.}{2014}]{Walsh2014}
{Walsh} A.~J.,  {Purcell} C.~R.,  {Longmore} S.~N.,  {Breen} S.~L.,  {Green}
  J.~A.,  {Harvey-Smith} L.,  {Jordan} C.~H.,   {Macpherson} C.,  2014, \mn@doi
  [\mnras] {10.1093/mnras/stu989}, \href
  {https://ui.adsabs.harvard.edu/abs/2014MNRAS.442.2240W} {442, 2240}

\makeatother
\end{thebibliography}

\bsp	
\label{lastpage}
\end{document}